# Electron transparent graphene windows for environmental scanning electron microscopy in liquids and dense gases

Short title: Graphene windows for wet SEM

**Joshua D. Stoll and Andrei Kolmakov[1]**

Department of Physics, Southern Illinois University, Carbondale IL, 62901 USA

**Abstract**

Due to its ultrahigh electron transmissivity in a wide electron energy range, molecular impermeability, high electrical conductivity and excellent mechanical stiffness the suspended graphene membranes appear to be a nearly ideal window material for *in situ* (*in vivo*) environmental electron microscopy of nano- and mesoscopic objects (including bio-medical samples) immersed in liquids and/or in dense gaseous media. In this communication, taking advantage of little modification of the graphene transfer protocol on to metallic and SiN supporting orifices, the reusable environmental cells with exchangeable graphene windows have been designed. Using colloidal gold nanoparticles (50 nm) dispersed in water as model objects for scanning electron microscopy in liquids, the different imaging conditions through graphene membrane have been tested. The limiting factors for electron microscopy in liquids such as electron beam induced water radiolysis and damage of graphene membrane at high electron doses were discussed.

---

[1] Corresponding author: akolmakov@physics.siu.edu; phone: +1-(618)-453-5212

1. Introduction

The recent implementation of aberration correction in scanning electron microscopy (SEM) resulted in demonstrated ability to achieve an atomic resolution using secondary electrons (Zhu *et al.*, 2009) what can potentially revolutionize the tooling used for imaging of the surfaces and interfaces. The next step will be the realization of this technique on the samples being under the real word environments. The traditional scanning electron microscopy, however, requires high vacuum conditions to avoid significant scattering of the primary (PE), secondary (SE) and backscattered (BS) electrons by residual gas or in the vicinity to the sample. The implementation of the differential pumping stages and novel gaseous electron detectors in SEM in early eighties (Danilatos, 1988) resulted in significant relief of the vacuum requirements and nowadays the routine imaging of the samples at few tens of Torr of unreactive gases *i.e.* water vapor can be performed using commercial environmental scanning electron microscopes (ESEM) (Donald, 2003). Truly *in situ* scanning electron microscopy and spectroscopy in electrolytes, water, reactive liquids and gases would provide a nanoscopic access to processes taking place at solid-liquid-gas interfaces what is strongly demanded by a variety of applications *i.e.* fuel cells, batteries, catalysis, (bio-) medical, automotive, geological, forensic *etc*. The latter however, remains a challenging task in electron microscopy. To address these needs a number of enclosed environmental cell (E-cell) designs have been developed since early forties to image the samples in liquids and or gases at (or near to) atmospheric pressure (Abrams and McBain, 1944; Swift and Brown, 1970; Parsons, 1974; Spivak *et al.*, 1977; Robinson, 1975) . These approaches employ few tens (up to few hundred) nanometers thin membranes which are sufficiently transparent for high energy (usually above 10 keV) primary and backscattered electrons to probe the interface or objects adhered (or placed in close proximity) to the inner side of the membrane. Few commercial E-cells designs (both: enclosed and flow-through) for *in situ* SEM and HRTEM are currently available (See www.2spi.com/catalog/grids/silicon-nitride-wet-cell-kits.php *et al.*) and a number of novel results or demonstrations have been reported lately (Williamson *et al.*, 2003; Gai *et al.*, 2008; Sharma, 2001; Jonge *et al.*, 2009; Thiberge *et al.*, 2004b; Thiberge *et al.*, 2004a; Grogan and

Bau, 2010; Kolmakova and Kolmakov, 2010; Donev and Hastings, 2009; Park *et al.*, 2012; White *et al.*, 2012; Liao *et al.*, 2012; Tai *et al.*, 2012; Wook Noh *et al.*, 2012; Xin and Zheng, 2012) (see also recent review (de Jonge and Ross, 2011) and references therein).

With recent advances in large scale fabrication of high quality single and multilayer graphene (and graphene derivatives) (Reina *et al.*, 2008; Li *et al.*, 2009a; Kim *et al.*, 2009; Bae *et al.*, 2010), its comprehensive characterization and development of high yield transfer methods (Aleman *et al.*, 2010; Liang *et al.*, 2011; Regan *et al.*, 2010) this 2D materials appear to be a new prospective windows platform for membrane based E-cells. In particular, graphene has a unique combination of properties excelling in mechanical strength (Lee *et al.*, 2008; Booth *et al.*, 2008), electron transparency (Meyer *et al.*, 2007; Meyer *et al.*, 2008; Muellerova *et al.*, 2010; Knox *et al.*, 2008) as well as being impermeable to liquids and gases (Bunch *et al.*, 2008; Stolyarova *et al.*, 2008; Xu *et al.*, 2010). Very recently the successful implementation of graphene and graphene oxide based enclosed liquid E-cells for SEM (Krueger *et al.*, 2011), HRTEM (Yuk *et al.*, 2012) and x-ray photoelectron microscopy have been demonstrated (Kolmakov *et al.*, 2011). Graphene oxide (GO) membranes (Dikin *et al.*, 2007) routinely made of overlapping GO sheets segregated from aqueous dispersions, as prepared, are water permeable (Nair *et al.*, 2012) and therefore require a special treatment to become liquid/vapor tight for extended period of time sufficient for SEM imaging (Krueger *et al.*, 2011). Therefore, the membranes made of high quality large domain graphen are most preferable for practical *in situ*

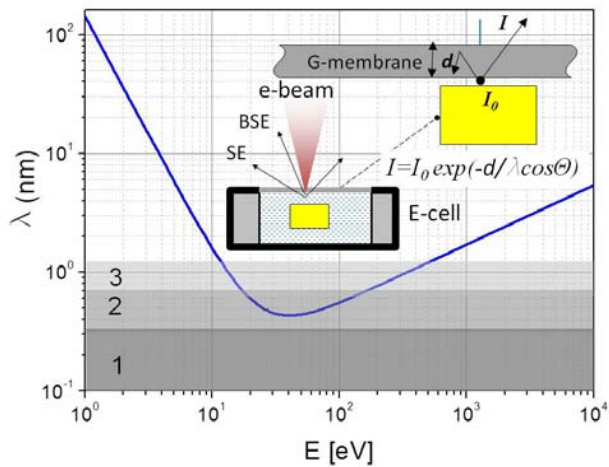

**Figure 1**. The dependence of inelastic mean free path $\lambda$ for electrons in carbon as a function of their kinetic energy calculated using NIST SRD-71 database [46]. Gray areas in the bottom represent a stack of three layers of graphene assuming 3.4 Å ML thickness. Inset: a diagram of the SEM experiment with graphene window based E-cell containing the sample immersed in dense media. The formula for attenuation of the outgoing electrons $I_0$ by the graphene membranes represents the case where the sample is adhered to the backside of the g-membrane

SEM application.

The general geometry and principle of the electron microscopy in liquid or dense gaseous environment using graphene E-cell are depicted in the Fig. 1. The primary electron beam (e-beam) penetrates through the g-membrane, interfacial liquid (gas) layer and impinges the sample below it thus creating outgoing flux of secondary and backscattered electrons from all these components. Provided that g-membrane is homogeneous, the contrast in the SEM image between the sample and liquid (or gaseous) background is a result of a difference in their SE and BSE emission yields and particularities of their scattering on their way to detector(s). In particular, the outgoing electrons get scattered by the interfacial layer (*i.e.* water) and by the membrane itself on their way to detector, thus the amount of electrons useful for imaging is reduced. The attenuation and scattering of the primary beam as well as outgoing BSE and especially SE by the mesoscopic layers of liquid or dense gas is very effective, thus the imaging of the object located few microns deep from the membrane is challenging under standard SEM operation conditions (Thiberge *et al.*, 2004b; Kolmakova and Kolmakov, 2010). For the practical reasons therefore, it is preferable when the sample is adhered or placed in close proximity (less than a 100 nm) to the inner side of the g-membrane. Different from commercial membranes made of SiN, SiC, $SiO_2$ and polyimide, SEM windows made of graphene have few important advantages such as: (i) low atomic number (low Z), what in combination with graphene's monolayer thickness, is beneficial for high electron transmissivity; (ii) high electrical conductivity which will eliminate parasitic charging effects and artifacts common for standard dielectric membranes; (iii) superior mechanical stiffness of graphene allows to withstand the pressure differential up to several Bars for membrane windows few microns wide (Lee *et al.*, 2008; Booth *et al.*, 2008). Since the attenuation of the primary electrons as well as BS and SE electrons depends on inelastic scattering, the advantages of the graphene as electron transparent membrane can be summarized in the (Fig.1) which depicts the energy dependence of the inelastic mean free path calculated for carbon using NIST SRD-71 algorithm (Jablonski and Powell, 2003, 2009). Based on these estimations, 0.34 nm thin carbon membrane does not scatter inelastically incoming and outgoing electrons and its transparency is only limited by elastic events. High transmissivity between 51% and 75% has been experimentally recorded for a suspended single graphene sheet for the electron energy 100 eV (Mutus *et al.*, 2011) and 73% for 66 eV electrons (Longchamp *et al.*, 2012). As also can be seen in the Fig. 1 even a three monolayer thick graphene (and therefore mechanically robust) membrane will be nearly transparent to electrons with kinetic energies (KE) above 400 eV and as low as 10 eV. The latter, opens very important imaging opportunities. Different form the standard 50-100 nm commercial membranes, which require high energy BSE for image formation, the transsmisivity of the graphene to low energy electrons opens an opportunity to employ SE instead of BSE for imaging of the objects behind the membrane. The latter would drastically increase the surface sensitivity and will reduce the energy dose deposited into the sample. In addition, the accessibility to this particular electron energy range, allows one to employ the unique capabilities of low energy electron point source (LEEPS) microscopy (Beyer and Goelzhaeuser, 2010), low energy electron microscopy (LEEM) (Bauer, 1994), photoelectron emission electron microscopy (PEEM) (Rotermund *et al.*, 1990) and scanning low energy electron microscopy (LESEM/SLEEM) (Frank *et al.*, 2011) in environmental research. In this communication, we report the first results on: (i) fabrication of graphene bases E-cells; (ii) imaging of fully hydrated model samples using standard SEM and (iii) revealing the possible artifacts as well as parasitic effects originating at high electron dose.

## 2. Experimental

The general design of the graphene E-cell is shown in the Fig. 2. Two kinds of graphene samples (both from Graphene Laboratories Inc**.**) were used in this study. CVD graphene on Cu foil was used to fabricate single layer graphene window (g-window) while multilayer graphene grown on Ni film on Si wafer was used to produce thicker and more robust windows. The commercial stainless steel orifices (Lenox Laser Inc. 10 mm in dia and 0.05-0.1 mm thick) having laser

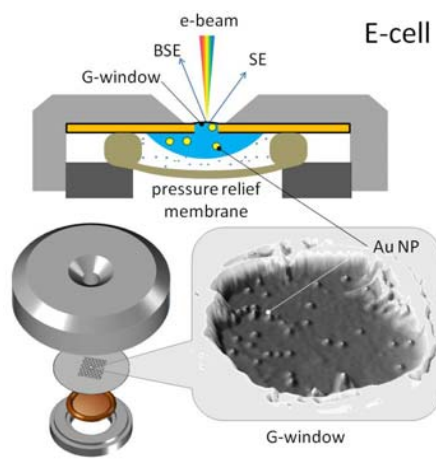

**Figure 2**. General design of g-window enclosed E-cell and a 3D presentation of the SEM image of g-window covering colloid 50 nm Au NP water solution. The image was taken at 20 keV using SED

drilled pinholes with diameters between 2 and 10 micron were used as exchangeable supporting frames for g-windows. Prior to graphene transfer, the orifices were electro polished in 45% phosphoric acid, 30% sulfuric acid, and 25% glycerol (by volume) solution to eliminate the laser induced edge roughness near the pinhole. To transfer graphene to these stainless frames we have used standard PMMA based protocol further upgraded in ref. (Li *et al.*, 2009b) with minor modifications. Briefly, after transfer of the PMMA film with the graphene layer from the distilled water onto the target substrate, the assembly was first dried in air and then was exposed to the saturated acetone vapor above the 65 °C hot acetone bath to induce gradual softening and melting of the polymer support. This procedure eliminates the effect of mesoscopic interfacial roughness of the contact between graphene and support and ensures tight adhesion of the graphene to the substrate. The polymer film softening and melting was visually monitored and controlled in real time with optical microscope. After the complete adhesion of melted polymer was achieved the entire assembly was slowly immersed in to the hot acetone bath to dissolve completely PMMA film. The second larger warm bath of acetone was used for final cleaning of the g-window from possible polymer residue. This additional cleaning of the g-window in a second hot acetone bath was found to be crucial in our design since the organic residue can be formed inside high aspect ratio few micrometer wide pinhole channel of the orifice upon drying of not sufficiently clean solvent. We found that entire protocol works as fine using drop casted simple nail polish lacquer diluted 1:10 in acetone instead of spin coated PMMA.

After the preparation and SEM inspection of the dry g-window, ca 10 µL droplet of uncojugated Au (50 nm) colloid, having passport density of $4.5 \cdot 10^{10}$ ml$^{-1}$ (BBI/Ted Pella), was placed on to the hydrophilic back side of the g-membrane using the micropipette. The E-cell was then quickly assembled and the droplet was sealed using o-ring/pressure relief membrane (Thiberge *et al.*, 2004b) (Fig. 2). Prior to imaging in SEM, the vacuum tightness of the E-cell and integrity of the g-window upon air evacuation was tested in a custom-made vacuum chamber interfaced with the optical microscope. Hitachi 4500-II FE SEM microscope with base vacuum $5 \cdot 10^{-6}$ Torr was equipped with: (i) in-lens (through-the-lens (TTL) or upper) secondary electron detector, (ii) lower SE detector (Everhart-Thornley (ET) type) and (iii) solid state 4-quadrant BSE detector placed above the sample, which were used for imaging with primary electron beam energy spanning the range from 2 keV to 30 keV and to test different contrast formation mechanisms. Monte Carlo (MC) simulations of the electron trajectories were performed using Casino v2.42 software (Hovongton *et al.*, 1996).

## 3. Results and discussion

Figure 3 gives comparative images of the same g-membrane a) before and b) after the cell was filled with liquid water. The same electron energy and detector (lower SE) was used in both experiments. In the left image the suspended graphene layer is barely visible at this energy due to high transmissivity for primary electrons as well as SE and BE electrons coming from the interior of the orifice. Assuming that the membrane is significantly transparent for primary 4 keV electrons, the effective tramissivity of this particular membrane for combined SE and BS electrons forming the image can be roughly estimated to be 80% comparing the gray scale values of the adjacent uncovered and membrane covered points (Fig. 3a). Filling the E-cell with the water drastically changes the observed membrane morphology and contrast distribution inside the graphene covered pinhole. The membrane's general shape becomes concave. The latter varies from the sample to sample and can be convex depending on the interplay between backing pressure and capillary forces. In addition, wrinkles appear due to roughness of the pinhole perimeter (Fig. 3 b). The ability to gain the topographic features from g-membrane indicates that contrast forming SE originate now not only from the membrane itself but also from the interfacial layer with the water. Different from the case of empty E-cell depicted on Fig. 3a, no any signal can be observed from the sidewalls. Instead, the filled membrane appears to be much darker compared to the Fig. 3a even in the enhanced brightness setting used in the Fig. 3 b. The latter is because the spurious SE and BS electrons emitted and/or scattered form the background sidewalls of the pinhole, become completely attenuated by liquid layer in the Fig. 3b, but are mainly responsible for contrast formation in Fig. 3a. This is supported by comparative MC simulations, shown in the Fig. 3c for the empty (left panel) and water filled space (right panel) between the membrane and the sidewall. The estimated electron range in the water (Joy and Joy, 2006) at 4 keV does not exceed one micron what implies that neither SE nor BS electrons from the metal support are able to contribute to the signal in a wet cell.

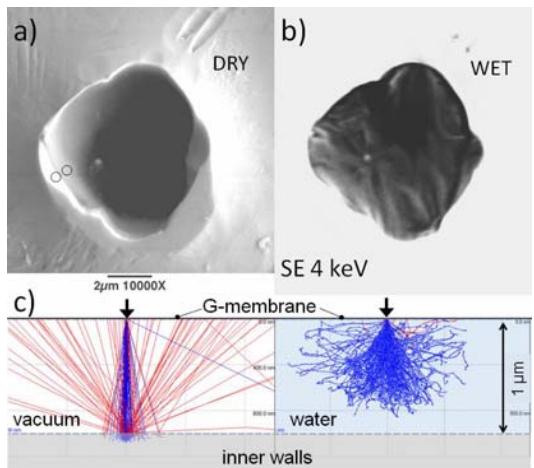

**Figure 3.** SEM images of the graphene membrane taken with ET type SE detector: a) in a dry state and b) under the wet conditions; c) MC simulations of the electron trajectories of the model stainless steel sample placed 1 micron behind graphene membrane. Blue (darker) trajectories correspond to the primary 4 keV electron beam and the red (lighter) ones are backscattered electrons; c) the same sample but with 1 micron thick intermediate water layer between the membrane and the substrate.

Additional insight on specificity of the electron interaction with g-membranes can be obtained via the direct comparison of images acquired from different electron detectors and at different energies of primary electrons. Figure 4a show the image of the ca 4 ML thick g-wingow of water filled E-cell and recorded with SE lower detector at 20 keV of the primary beam. The same g-window was imaged under the same conditions in the Fig. 4b but using BSE detector. The electron range in water at this energies is in the order of 10 microns (Joy and Joy, 2006) and as a result BS electron detector is capable to image the pinhole itself located deep under the water filled membrane. In addition, the diffuse halo around the pinhole can be observed using BSE, which is due to the interfacial water layer formed between the g-membrane and stainless steel substrate. Compared to the dark pinhole, larger electron signal can be detected in this area, since BS electrons from the background metal substrate are capable to penetrate through this interfacial water layer. The effective scattering of electrons on their way out of the substrate coupled with the variable thickness of interfacial water layer produces the smooth gradient of gradient og the gray scale as well deteriorates the resolution of the image and thus forms a halo. This is complemented by the image in the Fig. 4a recorded with SE (lower) E-T detector which has high sensitive to low energy secondary electrons. Different to prior case, the surface topography of the g-membrane can now be seen in details and the real size of the water-filled pocket under the membrane, that exceeds the diameter of the pinhole can be measured. In addition to reveling the real topography of the membrane and image forming mechanisms, these two complementing images demonstrate the water being able to weaken the interaction between graphene layer and the substrate and "creep" under the transferred graphene membrane as a result of pressure differential. Though the mechanical stiffness of 4 ML membrane is sufficient to sustain the pressure differential, the detailed comparison of higher magnification images in the panels c and d reveals the appearance of dark dendritic features visible in the top of the membrane in the Fig. 4 c. These are water-filled wrinkles of the graphene sheet, which serve as primary spreading channels of the interfacial water. Figure 4 c and d show zoomed area of the g-membrane imaged with in-lens and lower SE detectors taken at 4 keV. Both SE detectors are sensitive to the surface topography however, the image taken with in-lens detector exhibits significantly reduced contrast between metallic support and g-membrane. This can be explained by different contribution of BS electrons in contrast formation in these two detectors. E-T (lower) detector provides significant Z contrast directly due to off-axis BSE and indirectly via detection of BSE induced $S_2$ and $S_3$ secondary electrons (Goldstein, 2003). On the other hand, BSE contribution is largely filtered out in TTL SE detector. As a result, low Z g-membrane/water filled area appears darker in Fig. 4c compared to Fe, Ni rich support which has significantly larger Z and therefore higher back scatter coefficient.

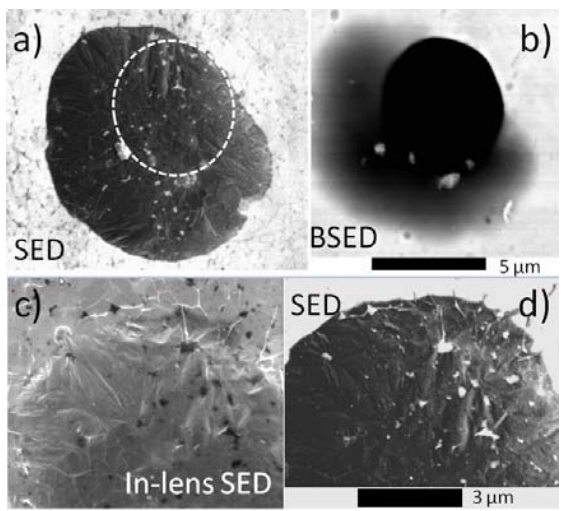

**Figure 4.** Water filled E-cell with 4 ML graphene membrane imaged with: a) E-T lower detector $E_p$=20 keV ; b) BS electron detector $E_p$=20 keV; c) TTL (upper) SE detector $E_p$=4 keV and d) E-T (lower) detector $E_p$=4 keV

Figure 5 exemplifies SEM images of 50 nm colloid Au nanoparticles dissolved in water and recorded at relatively low (Fig. 5 a; 2 keV) and high (Fig. 5 b-e; 20 keV) electron beam energies correspondingly. Remarkably, the electron transparency of g-membrane with nominal thickness of 4 ML is sufficient to be able to observe Au nanoparticles using low energy SE sensitive TTL detector and as low as 2 keV primary beam energy (Fig. 5a). For evaluation of the g-membrane's transmissivity, the series of the images of the membrane was recorded as a function of energy using the same detector and average gray scale values $S_{Au}$ of the individual Au nanoparticles as well as neighboring background values $S_{BG}$ have been recorded. The contrast $C= (S_{Au}-S_{BG})/S_{Au} = 0$ would indicate that g-membrane is not transparent while the values $C > 0$ indicate that it is transparent enough to be used for ESEM.

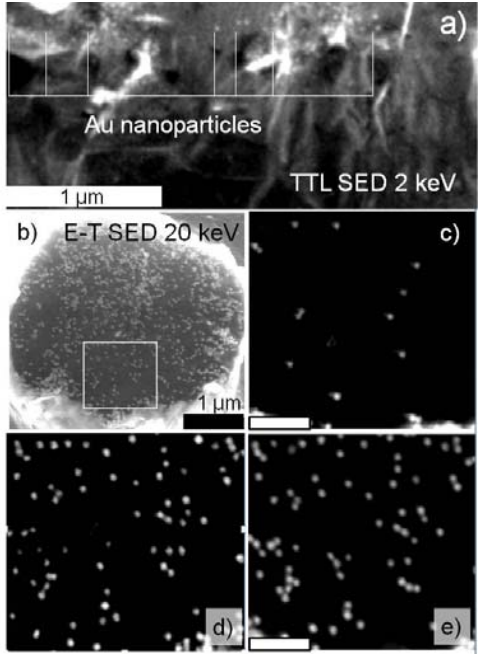

**Figure 5**
a) Au nanoparticles adhered to the backside of 4 ML g-membrane imaged with TTL SE detector and $E_p$=2 keV; b) typical SEM image of E-cell filled with Au 50 nm colloid nanoparticles in water. Image conditions: $E_p$=20 keV, E-T SE detector; Higher magnification SEM images of the same g-membrane taken at: c) 30 minutes d) 225 and e) 242 minutes after the E-cell was filled with colloid NPs

The $C$ values of 0.43±0.4, 0.6±0.12 and 0.54±0.08 have been measured for the g-membrane with nominal thickness 4 ML for 2, 4 and 20 keV primary electron beam correspondingly using TTL SE detector. Compared to commercial SiN, $SiO_2$ and polyimide membranes (Ouantomix) as well as to 20-40 nm thick GO membranes (Krueger *et al.*, 2011), these values show that few ML thin g-membranes are significantly more transparent at low energy domain. The work is in progress to examine the transparency of the g-membrane at ultra-low electron energies.

*In situ* SEM/STEM imaging in liquids allows observing variety of dynamic processes such as nanoparticle and nanowire growth (Radisic *et al.*, 2006; Kolmakova and Kolmakov, 2010; Yuk *et al.*, 2012; Zheng *et al.*, 2009), their diffusion (Zheng *et al.*, 2009; de Jonge and Ross, 2011), electron beam induced nanoparticle mobility (Krueger *et al.*, 2011; White *et al.*, 2012) and dissolution/deposition (Donev and Hastings, 2009; Wook Noh *et al.*, 2012) from solution and *etc* which proceed slower, compared to beam rastering rate. Figures 5 b show the typical SEM image of Au colloid solution with concentration of ca $4.5 \cdot 10^{10}$ ml$^{-1}$ taken through the g-membrane more than 200 minutes after the filling of the E-cell. Similar to standard *in situ* electron microscopy studies using SiN TEM windows, only those nanoparticles, which adhere to the backside of the g-membrane, can be imaged with SEM reliably. Imaging these membranes as a function of time reveals that initial surface density of the adhered NP in the field of view (FOV) is close zero (not shown here) what agrees with high dilution level ($4.5 \cdot 10^{10}$ ml$^{-1}$) of NP in water and therefore low initial number of NPs in few microns wide and tens micron long orifice channel. Due to the Brownian diffusion of the nanoparticles in the channel, the nanoparticles eventually adhere to the back-side of the membrane as well as to channel's walls) and thus can be imaged. Figures 5 c-e were acquired sequentially from the area marked with white square in the Fig. 5b starting from *ca* 30, 225, and 242 minutes from the loading the cell with Au colloid solution. As predicted, the amount of the adhered NP in the FOV increases with time until complete saturation. The latter is determined either by established kinetic equilibrium between NP attachment/detachment rates or by complete precipitation of nanoparticles from solution to the membrane, channel walls and substrate surface. To evaluate this NPs transport scenario, the diffusion time inside the channel was estimated via classical mean-square displacement $\tau = \alpha <\chi^2>/2kT$ where $<\chi^2>$ is a mean-square NP displacement, which in our case would be equal the length of the channel, $\alpha = 6\pi\eta\delta$ is a drag coefficient, $\eta \sim 10^{-3}$ N·s/m$^2$ is an absolute viscosity of water at room temperature and $\delta$=50 nm is NP's diameter. The estimated diffusion time was found to be in the order of tens minutes what corroborates with observed relatively long "appearance" time of NPs in the FOV. The steady state surface density of the nanoparticles gained in the images 5e and 5d, is lower compared to the one can expect upon complete precipitation of NPs from the solution. The latter can be explained by the fact that the observed solution is not completely dried yet and/or relatively weak adhesion of NPs to the g-membrane.

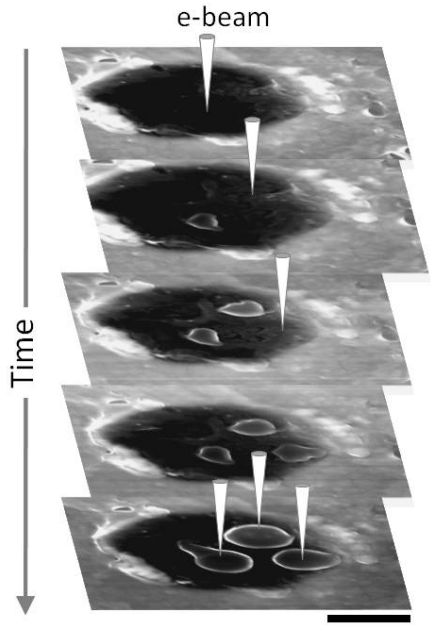

**Figure 6**
The formation of the hydrogen containing bubbles under the g-membrane because of beam induced water radiolysis. The scale bar corresponds to 1 micron. Beam energy 20 keV, electron current $10^{-10}$ A

The description and performance of the SEM environmental cells equipped with ultra thin g-membranes would not be complete without the brief discussion of the observed limitations due to radiation damage, which is particularly important for the samples immersed in the dense gaseous or liquid media (Royall *et al.*, 2001). The enhanced radiation damage of soft and biological matter under hydrated conditions by focused electron beams have been thoroughly discussed for environmental SEM as well as for E-cell based SEM (Danilatos, 1981; Kirk *et al.*, 2009; Thiberge *et al.*, 2004a). The reverse effect of damaging the membranes themselves via electron-induced effects in the liquid media is less studied, however is crucial for defining the ultimate magnification and operation time which can be achieved before the radiation damage deteriorates the membrane and/or the sample. We have observed irreversible changes in the g-membranes morphology and their disruptions when the electron dose exceeds the threshold of $10^4$ e/nm$^2$. Figure 6 (top) shows the typical multilayer g-membrane filled with water. The electron beam with energy 20 keV and a current ~$10^{-10}$ A was focused on the area ca 250x250 nm and images were acquired with TV scanning rate (second panel of the Fig.6). The radiation damage usually originates within ca 10 seconds of scanning and visible bubble is formed under the g-membrane in the exposed area (see supporting material). The procedure is reproducible at any other location of the g-membrane. Zooming out the area halts the bubble formation, while the extended exposure leads to growth of the bubbles (see bottom panel in the Figure 6) and eventually to g-membrane disruption. The similar beam induced effects were routinely observed during high dose irradiation of liquid (Thiberge *et al.*, 2004b) or frozen hydrated samples (Leapman and Sun, 1995) and manifest the effect of beam induced radiolysis of water. This process was thoroughly studied (see as an example ref. (Royall *et al.*, 2001) and references therein) and briefly can be described as follows. Upon dissipation of energy of the electron beam in water, a variety ionized and exited molecular species are created in the interaction volume. Most of them recombine rapidly but few chemically reactive products such as molecular hydrogen ($H_2$), hydrogen peroxide ($H_2O_2$) and hydroxyl radical ($\cdot OH$) accumulate due to ether chemical stability or lack of the effective decay channels. As a result, hydrogen containing bubbles can be formed under the membrane which inner pressure can be well in access of several Bars. The latter unavoidably will lead to the membrane disruption. In addition, in spite of the demonstrated chemical stability of the graphene (Chen *et al.*, 2011), the hydrogen peroxide and hydroxyl radical can oxidize g-membrane along the grain boundaries and at point defects what again will lead to destruction of pressurized membrane. The E-cells used in this study were designed to be for a single use. However, the good practice for imaging of fully hydrated samples for extended period would be not exceeding this "radiolysis dose threshold".

## 4. Summary

In conclusion, we proposed and tested the reusable enclosed E-cell design for environmental SEM with exchangeable electron transparent windows made of single and multi layer graphene. The electron transparency of the g-membrane was evaluated for 2, 4 and 20 keV using Au colloid nanoparticles as a model of fully hydrated nanobjects. Using a number of different electron detectors, the different contrast formation mechanisms of water filled g-membranes have been discussed. Similar to commercial membranes, the chemical interaction of the g-membrane with reactive species, formed because of water radiolysis at high electron doses, reduces the lifetime of this prospective window material.

**Acknowledgements**

The research was conducted on Hitachi 4500 FE SEM, which was generously donated to AK's lab by NIST. The support with samples from Lenox Laser and TEMwindows.com a division of SiMPore Inc. is greatly appreciated. The authors are thankful to Dr. David C. Joy (ORNL), Prof. John Bozzola for encouraging discussions as well as to Mr. Clay Watts and Mr. Robert Bayer (all at SIUC) for technical and computer support. Many thanks to Ms. Jie Zhang for preparation of the graphic material for this manuscript. The research was supported through NSF ECCS-0925837 grant.